\RequirePackage{fix-cm}
\documentclass[twocolumn]{svjour3}          
\smartqed  
\usepackage{graphicx}
\usepackage{array}
\usepackage{amssymb}
\usepackage{wrapfig} 
\usepackage{enumerate}
\usepackage[shortlabels]{enumitem}
\usepackage{pifont}
\usepackage{cite}
\usepackage{booktabs}
\usepackage{amsmath}

\usepackage{enumitem}

\usepackage{caption}
\usepackage{subcaption}
\usepackage{tgtermes} 

\begin{document}
\sloppy

\title{A Method for Analyzing Stakeholders' Influence on an Open Source Software Ecosystem's Requirements Engineering Process}


\author{
        Johan Lin{\aa}ker \and 
        Bj{\"o}rn Regnell \and 
        Daniela Damian
}


\institute{
            J. Lin{\aa}ker \at
              Lund University, Box 118, SE-221 00 Lund, Sweden \\
              Tel.: +46 46 222 49 27\\
              \email{Johan.Linaker@cs.lth.se}           
          \and
          B. Regnell \at
              Lund University, Box 118, SE-221 00 Lund, Sweden \\
              \email{Bjorn.Regnell@cs.lth.se}
          \and
          D. Damian \at
              University of Victoria,
              PO Box 1700, STN CSC Victoria, BC, Canada \\
              \email{Damian.Daniela@gmail.com}
}

\date{Received: date / Accepted: date}

\maketitle

\begin{abstract}
\textbf{Background:} For a firm in an Open Source Software (OSS) ecosystem, the Requirements Engineering (RE) process is rather multifaceted. Apart from its typical RE process, there is a competing process, \emph{external} to the firm and inherent to the firm's ecosystem. 
When trying to impose an agenda in competition with other firms', and aiming to align internal product planning with the ecosystem's RE process, firms need to consider who and how influential the other stakeholders are, and what their agendas are.
\textbf{Aim:} The aim of the presented research is to help firms identify and analyze stakeholders in OSS ecosystems, in terms of their influence and interactions, to create awareness of their agendas, their collaborators, and how they invest their resources.
\textbf{Method:} To arrive at a solution artifact we applied a design science research approach where we base artifact design on literature and earlier work.
\textbf{Results:} A Stakeholder Influence Analysis (SIA) method is proposed and demonstrated in terms of applicability and utility through a case study on the Apache Hadoop OSS ecosystem. SIA uses social network constructs to measure the stakeholders' influence and interactions and considers the special characteristics of OSS RE to help firms structure their stakeholder analysis processes in relation to an OSS ecosystem. 
\textbf{Conclusions:} SIA adds a strategic aspect to the stakeholder analysis process by addressing the concepts of influence and interactions, which are important to consider while acting in collaborative and meritocratic RE cultures of OSS ecosystems.

\end{abstract}

\section{Introduction}
\label{sec:intro}


Firms that use Open Source Software (OSS), e.g., as part of their supporting infrastructure, product strategy or business model, need to consider the Requirements Engineering process of the OSS itself~\cite{munir2015open}. This second, \emph{external} to the focal firm, RE process is facilitated by the software ecosystem (cf. OSS community~\cite{nakakoji2002evolution}) that surrounds the OSS~\cite{jansen2009business}. Firms that are users of the OSS may also be involved in its development and maintenance and can be considered as members of the ecosystem, as well as stakeholders to the OSS. We refer to Glinz \& Wieringa's definition of a stakeholder as \textit{``\ldots a person or organization who influences a system's requirements or who is impacted by that system''}~\cite{glinz2007guest}. In our context, we consider a person or an organization as the members of an OSS ecosystem, and the system being the OSS that underpins the ecosystem, using the definition by Jansen et al~\cite{jansen2009business}. 

RE practices in OSS ecosystem may be described as informal and decentralized. There is often no central repository with requirements defined in the problem space, describing the product of need, along with heavy processes and tools for examining the requirements for completeness and consistency~\cite{alspaugh2013ongoing}. Instead, RE may be considered as a lightweight and evolutionary process of requirements refinement~\cite{ernst2012case}. Practices such as elicitation, specification, and prioritization overlap and are done collaboratively through iterative and transparent discussions and negotiations including up-front implementations~\cite{scacchi2002understanding, german2003gnome, ernst2012case}. These discussions and implementations of requirements are spread out over a number of requirements artifacts, each with its own repository. Examples of these artifacts (cf. informalisms~\cite{scacchi2002understanding}) include reports in an issue tracker, messages in a mailing list, or commits in a version control system. Prioritization is commonly conducted by stakeholders with central positions in the ecosystem's governance structure~\cite{laurent2009lessons, baars2012framework}. To gain such a position in OSS ecosystems with a meritocratic governance structure, a stakeholder needs to prove merit by being active, contributing back, and having a symbiotic relationship with the OSS ecosystem~\cite{dahlander2005relationships}.


Hence, the focal firm is one stakeholder among many within an open and fluctuating population in the OSS ecosystem~\cite{jensen2007role}. This can result in conflicting agendas and lack of control, e.g., in regards to which requirements to be implemented and prioritized, render misalignment with internal RE processes~\cite{wnuk2012can}, and complicate contribution strategies~\cite{munir2015open}. The focal firm may, therefore, have to gain the influence necessary to affect the RE process in an OSS ecosystem according to its own agenda.

The Merriam-Webster dictionary~\footnote{http://www.merriam-webster.com/dictionary/influence} defines influence as \textit{``the power to change or affect someone or something''}. In our context, this relates to the power of a stakeholder to change or affect the RE process in an OSS ecosystem. This notion of influence aligns naturally with what defines a stakeholder~\cite{glinz2007guest}, and as a characteristic enables firms to, e.g., see the requirements in which stakeholders hold a certain interest, and from there be able to create an overview of their agendas in the ecosystem~\cite{frooman1999stakeholder}.
Further, this understanding enables the focal firm to analyze how these stakeholders invest their resources in order to satisfy their agendas~\cite{frooman1999stakeholder}. By also considering other stakeholders' interactions within the ecosystem, firms may identify possible partners and competitors~\cite{rowley1997moving}. Moreover, this can help firms to learn how to adapt their own strategies and processes with the OSS ecosystem's and how to build their own influence and position the ecosystem's governance structure~\cite{baars2012framework}. 
The knowledge output can then be leveraged towards other stakeholders through the politics and negotiations that take place in the ecosystem's RE process~\cite{Milne2012}.

These aspects highlight the importance of stakeholder identification and analysis as input to the continuous and complex decision-making process which RE constitutes~\cite{aurum2003fundamental} by helping to answer questions as which other stakeholders exist in the ecosystem, what are their agendas, and how do they aim to achieve them~\cite{frooman1999stakeholder}. 
However, current practices~\cite{pacheco2012systematic} are not adapted to consider these strategic aspects~\cite{freeman2010strategic} in the context of OSS ecosystem~\cite{munir2015open} and its informal and collaborative RE process~\cite{scacchi2002understanding, ernst2012case}, specifically the importance of understanding stakeholders' influence and interactions. Involved firms are no longer the vantage point, and instead, form part of a larger set of interdependent stakeholders~\cite{rowley1997moving}.
We address this gap with a design science research approach~\cite{wieringa2014design, hevner2004design} and define it as a design problem~\cite{wieringa2014design}:

\begin{enumerate}
    \item[\textbf{DP}] \textit{How to characterize the influence of stakeholders on the OSS ecosystem's RE process, so that a focal firm can understand other stakeholders' agendas, collaborations, and resource investments in pursuing these agendas?}
\end{enumerate} 

The contribution of our work is the proposal of the Stakeholder Influence Analysis (SIA) method. Its aim is to help firms to analyze an OSS ecosystem to identify its stakeholders' influence by the impact they have with respect to the requirements that get implemented in the OSS. We base SIA on social network analysis constructs~\cite{wasserman1994social, faust1997centrality, newman2010networks} that have proven to be useful in characterizing the influence of stakeholders~\cite{rowley1997moving, orucevic2014network}, but also effective when analyzing a firm's participation in OSS ecosystems~\cite{orucevic2014network, teixeira2015lessons} and requirement-centric stakeholder collaborations~\cite{damian2007collaboration, marczak2008information, bhowmik2015on}. An analysis approach used in an earlier reported case study of the Apache Hadoop OSS ecosystem~\cite{linaaker2016firms} is formalized to consider how requirements may be informally represented in multiple artifacts in decentralized repositories present in OSS ecosystems~\cite{scacchi2002understanding, ernst2012case}. The influence analysis is then operationalized with a stakeholder mapping approach based on earlier work~\cite{johnson2008exploring, newcombe2003client, mendelow1991stakeholder}. To demonstrate SIA's applicability and utility, we present a case study of the Apache Hadoop OSS ecosystem.


The rest of this paper is structured as follows: In section~\ref{sec:researchApproach} we describe the research approach used in the development of SIA. In section~\ref{sec:SIA} we give a detailed presentation of SIA, while in section~\ref{sec:casestudy} we demonstrate its applicability and utility with a case study. In section~\ref{sec:discussion} we discuss alternative approaches to characterizing influence and threats to validity. Finally, we conclude the paper in section~\ref{sec:conclusions}.

\section{Research Approach}
\label{sec:researchApproach}


To develop SIA, we used a design science research approach~\cite{wieringa2014design, hevner2004design} where research is conducted iteratively through design cycles. A design cycle consists of three phases: problem investigation, artifact design, and artifact validation~\cite{wieringa2014design}. Below we describe these steps in detail.

\textbf{Problem Investigation phase:} Here, the research goal and the problem context are (re-)analyzed before any artifact is designed, or any improvements implemented~\cite{wieringa2014design}. In previous work~\cite{linaaker2016firms}, we explored how centrality measures could be used to characterize the influence of stakeholders within an OSS ecosystem, and how this evolved over time. Findings helped to create an understanding of the problem context and helped define the design problem (\textbf{DP}) as stated in Section~\ref{sec:intro}. In order to further understand the problem context, a literature survey was conducted to identify related work on:

\begin{itemize}
    \item the informal and collaborative RE processes within OSS ecosystems (e.g.,~\cite{scacchi2002understanding, ernst2012case, nakakoji2002evolution, jensen2007role, laurent2009lessons}),
    \item how awareness of the dynamics behind stakeholder interactions and interrelationships may be used to analyze their agendas (e.g.,~\cite{munir2015open, munir2017open, mitchell1997toward, frooman1999stakeholder, baars2012framework, rowley1997moving, pacheco2012systematic}), and
    \item how social network constructs may be used to characterize the stakeholders' interactions and influence on the RE process of the OSS ecosystem (e.g.,~\cite{wasserman1994social, faust1997centrality, newman2010networks, rowley1997moving, barnett2011encyclopedia, damian2010requirements, marczak2008information, bhowmik2015on, teixeira2015lessons, orucevic2014network}).
\end{itemize}

Surveyed literature provided conceptual foundations, which together with findings from previous work~\cite{linaaker2016firms}, constituted a knowledge base for the artifact design process.

\textbf{Artifact Design phase:} Here, knowledge gained from the previous phase is used as input to the design of an artifact with the hypothesis that it may act as a treatment for the design problem~\cite{wieringa2014design}. The Stakeholder Influence Analysis (SIA) method was formalized and structured as seven steps, as presented in Section~\ref{sec:SIA} (S1-S7) and in Fig~\ref{fig:SIA_Overview}. S1-S2 involves setting the purpose and scope of the analysis. S3 concerns data gathering, while S4-S6 concerns data processing. Finally, S7 regards the analysis of the processed data.

\textbf{Artifact Validation phase:} Here, the previously designed artifact is tested in the problem context in order to evaluate its treatment of the design problem~\cite{wieringa2014design}. To test SIA, we apply it in a proof of concept demonstration that it is functional and practical, 
through a case study on the Apache Hadoop OSS ecosystem (see Section~\ref{sec:casestudy}). It can be seen as an early form of descriptive validation where information from the knowledge base, and detailed scenarios can be used to demonstrate an artifact's applicability and utility~\cite{hevner2004design}. The Apache Hadoop OSS ecosystem was chosen due to the high concentration of firms in the ecosystem, and because it is the Apache project with the highest number of committers~\footnote{https://projects.apache.org/projects.html?number}. The case study further helped to evolve and refine SIA and its seven steps as can be expected by an iterative design process. 
\section{The Stakeholder Influence Analysis (SIA) method}
\label{sec:SIA}

\begin{figure*}[t!]
\centering
\includegraphics[scale=0.42]{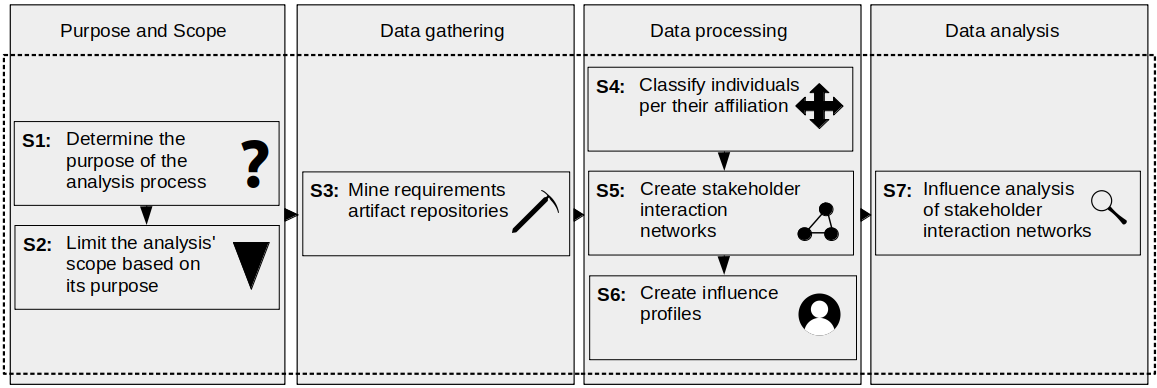}
\caption{Overview of SIA's seven steps (S1-S7) divided in Purpose and Scope, Data gathering, processing and analysis.}
\label{fig:SIA_Overview}
\end{figure*}

SIA aims to help firms involved in OSS ecosystems to structure their stakeholder identification and analysis process systematically when bridging their internal RE process with that of the ecosystem's (see Fig.~\ref{fig:SIA_Overview}). The focus is specifically on identifying and characterizing stakeholders' interactions and influence on the RE process in the OSS ecosystem. As proposed by Glinz and Wieringa~\cite{glinz2007guest}, SIA considers both individuals and organizations as stakeholders but primarily from an organizational level, meaning that the individuals in an OSS ecosystem should be aggregated to their organizational affiliation as far as possible. Below, we give a detailed overview of SIA and its seven steps, as outlined in Fig.~\ref{fig:SIA_Overview} and table~\ref{tbl:SIA}.

\begin{table*}[htbp]
\centering
\caption{Overview of SIA and its seven sequential steps (S1-S7) along with related descriptions and examples.}
\label{tbl:SIA}
\begin{tabular}{p{1cm}p{4cm}p{11cm}}
\toprule
& Step 
& Description 
\\ \midrule
\textbf{S1} 
& Determine the purpose of the analysis process. 
& Purpose could include:
\begin{itemize}
    \item Understand how an ecosystem is set up in terms of power-structure and general collaboration patterns.
    \item Identify potential partners or competitors as input to contribution strategies or collaborations.
    \item Identify stakeholders with aligning or conflicting agendas in regard RE-related activities and negotiations.
    \item Identify influential stakeholders to learn from in order to raise one's own influence in the OSS ecosystem.
\end{itemize}
\\
\textbf{S2} 
& Limit the analysis' scope based on its purpose. 
& Regards boundaries for what data that should be collected and is determined by the purpose of the analysis process. E.g., is the interest limited to:
\begin{itemize}
    \item a certain component or set of features of the OSS?
    \item a certain individual or set of stakeholders?
    \item a certain time-period or set of releases? 
\end{itemize}
\\
\textbf{S3} 
& Mine requirements artifact repositories. 
& Refers to the main repositories through which stakeholders interact in regards to the RE process. E.g., 
\begin{itemize}
    \item IRC or other chat-based communication
    \item Issue trackers
    \item Code review
    \item Software code repository
    \item Discussion boards
\end{itemize}
\\
\textbf{S4} 
& Classify individuals per their affiliation.
& Concerns identification of organizations to which individual developers are affiliated. E.g., by
\begin{itemize}
    \item Interacting and studying the communication within an OSS ecosystem.
    \item E-mail domain analysis.
    \item Heuristically through social media and public electronic sources.
    \item Identity pattern matching.
\end{itemize}
If no affiliation can be found or exists, the individuals can either be considered either as individual stakeholders or as an aggregated group.  
\\
&&\\
\textbf{S5} 
& Create stakeholder interaction networks. 
& For each requirement artifact repository, a directed and weighted affiliation-network is created. Stakeholders are represented as nodes, and are connected by edges if they have interacted on a common requirements artifact, e.g., commented on the same issue or mail-thread. To reflect investment and influence, edges are weighted based on the size of each stakeholder's participation.
\\
&&\\
\textbf{S6} 
& Create influence profiles. 
& To characterize stakeholders' influence on the RE process in the OSS ecosystem, a set of network centrality measures are calculated based on the interaction networks, and used to create an overall influence score. Together, they form an influence profile for each stakeholder. The centrality measures include:
\begin{itemize}
    \item Out-degree centrality
    \item Betweenness centrality
    \item Closeness centrality
    \item Eigenvector centrality
\end{itemize}
\\
&&\\
\textbf{S7} 
& Influence analysis of stakeholder interaction networks. 
& Based on influence profiles, stakeholders are ranked on overall influence score, and cross-compared on the centrality measures. Stakeholders of special interest are investigated further in regards to their relationships. With qualitative analysis of stakeholders' agenda alignment with the focal firm's, stakeholder mapping can be used with the influence/alignment matrix. The analysis should be directed by the purpose defined in \textbf{S1}.
\\
&&\\ \bottomrule
\end{tabular}
\end{table*}

\textbf{Determine the purpose of the analysis process (S1):}
The first step 
is to determine what questions are of interest to answer based on the stakeholder analysis. E.g., to identify potential partnerships or competitors, to identify and learn from stakeholders in a certain position, or to identify conflicting agendas in regards to certain requirements.

\textbf{Limit the analysis' scope based on its purpose (S2):}
Based on the purpose of the analysis process, limitations may be implied that can affect how the analysis should be narrowed down in terms of what requirements artifacts should be included in the analysis. E.g., is the analysis limited to:
\begin{itemize}
    \item a certain component or set of features of the OSS?
    \item a certain individual or set of stakeholders?
    \item a certain time-period or set of releases? 
\end{itemize}

\textbf{Mine requirements artifact repositories (S3):}
In the third step, the goal is \textit{to identify and mine the repositories that are mainly used by the OSS ecosystem}. Examples include issue trackers, mailing-lists, IRC logs, source code repositories, and code-reviews~\cite{scacchi2002understanding, ernst2012case}. When these are identified, the repositories should be mined to collect the necessary data. This can either be done either manually or with the help of existing\footnote{See e.g., https://metricsgrimoire.github.io/} or custom-made tools.

\textbf{Classify individuals per their affiliation (S4):}
In the fourth step, the \textit{individuals that are involved in OSS ecosystem need to be classified in regards to their affiliation}.
This is a necessary step as firm-affiliated individuals may be assumed to represent the agenda of their sponsor or employer~\cite{henkel2008champions, dahlander2006man}. However, not all individuals involved in an OSS ecosystem have to be affiliated and may rather represent their own personal agenda. These affiliations can be identified and triangulated by qualitative and quantitative means. E.g., through involvement and discussions, and by analyzing meta-data from the requirements artifact repositories and cross-checking against other information sources (e.g., social media and electronic archives)~\cite{bird2012examining, gonzalez2013understanding, linaaker2016firms}. 

If no affiliation can be found or exists, the individuals can either be considered as individual stakeholders or as an aggregated group. \textit{For example}, say, John, Mark, Lucy, Kate, and Mary are involved in the Apache Hadoop OSS ecosystem as developers. John and Kate work for a firm called Hortonworks and therefore have a common agenda. They are therefore aggregated and viewed as one stakeholder represented by the firm Hortonworks. Mark, Lucy, and Mary are all independent with the difference that Lucy is a relatively active user in the ecosystem, while Mark and Mary are more involved on a hobby basis. Lucy could, therefore, be seen as an independent stakeholder, while Mark and Mary could be aggregated to one group of hobbyists and be considered as one stakeholder. This type of classification and separation is rather subjective and needs to be done on a case-by-case basis for each ecosystem.

\textbf{Create stakeholder interaction networks (S5):}
In the following step, \textit{an interaction network for each requirements artifact repository needs to be created} in order to visualize the interactions between stakeholders. To create these networks, the interactions between the stakeholders to the requirement artifacts within a requirements artifact repository must be identified. As an example, consider a number of individuals (stakeholders) that discuss the need for as well as potential implementations of a new feature in an OSS project. The feature request is represented by an issue (requirements artifact) on the OSS ecosystem's issue tracker (requirements artifact repository). The discussions (interactions) between the individuals concerning the feature's evolution and refinement is recorded and persisted in the issue. This continuous discussion may be referred to as an ''event'' in social network theory~\cite{wasserman1994social}. The individuals partaking in the discussions may be referred to as ''participants'' of the same event~\cite{wasserman1994social}.

These events and their participants can furthermore be represented by networks of actors. Two actors within a network are connected by an edge if they have participated in the same event (as a network may include several events). If a network was created based on the previous example, all individuals who partook in the discussion of the issue would be represented by an actor in a network with an edge connecting each one of them. If there was a related discussion of the feature on the OSS ecosystem's mailing-list, a similar network may be created based on the concerned mail-thread. The two networks could then be analyzed in conjunction to get a more complete overview of the stakeholders to the requirement and their interactions (cf. requirement-central networks~\cite{damian2007collaboration}).

In a similar fashion, sets of requirements may be analyzed by aggregating requirements artifacts in a repository to a network. Returning to the example, a network could be created that included all of the issues in the issue tracker that are related to a certain release, created in a certain time span, or belonging to the same sub-module. A corresponding network could be created based on the mailing-list given that the same conditions apply. By creating corresponding networks of all the relevant requirements artifact repositories, the analyst may get a complete overview of what stakeholders that are involved and how they interact.


It should be noted that one stakeholder's participation in the event (e.g., RE-related discussions of an issue) may be of a relatively different size than the other stakeholders'. A stakeholder with a higher degree of participation may be considered to have a larger investment and interest in the event. These differences in the investment of time and resources need to be considered in order to give a fair view of a stakeholder's stake in a requirement. The relative size of the investment also helps to give a fairer data-set when doing an influence analysis of the interaction networks. As suggested by Orucevic-Alagic et al.~\cite{orucevic2014network}, weights can be calculated to describe the relative size of the participation to an event. 


Following Orucevic-Alagic et al.~\cite{orucevic2014network}, for a set of stakeholders \(V = \{v_1, v_2, ..., v_k\} \) and a set of requirements artifacts (events) \(U = \{u_1, u_2, ..., u_m \} \), we define a weight \(W\) of an edge between one stakeholder \(v_i\) and all other stakeholders that collaborate on an artifact \(u_t\) as:

\[ W(v_i, u_t) = \frac{X(v_i, u_t)}{\sum_{c=1}^k X(v_c, u_t)} \]

where \(X(v_i, u_t\)) denotes the number of times a stakeholder \(v_i\) has participated  in the collaboration on the requirements artifact \(u_t\).

Continuing from Orucevic-Alagic et al.~\cite{orucevic2014network}, this means that the weight of the edge \(W(v_i, v_j)\) for all requirements artifacts that two stakeholders \(v_i\) and \(v_j\) have collaborated on together equals:

\[ W(v_i, v_j) = \sum_{t=1}^m W(v_i, v_j, u_t) \]

As an example, when creating an interaction network based on an issue-tracker, each issue represents a requirements artifact and number of posted comments may represent the size of participation (\(X\)) of a stakeholder. Given that three stakeholders \(v_A\), \(v_B\) and \(v_C\) comment on the issue, they are all considered as actors in a network with edges connecting them. The weights would, therefore, consider the relative number of comments of each stakeholder as the size of their participation. Say \(v_A\) commented 1, \(v_B\) commented 2, and \(v_C\) commented 3 times. This results in the edge weights: 

\begin{itemize}
    \item \(W(v_A,v_B) \& W(v_A,v_C)\ = 1/5\)
    \item \(W(v_B,v_A) \& W(v_B,v_C)\ = 2/5\) 
    \item\(W(v_C,v_A) \& W(v_C,v_B)\ = 3/5\)
\end{itemize}

\begin{figure}[]
\centering
\includegraphics[scale=0.45]{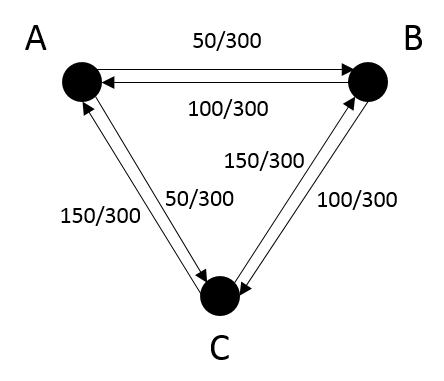}
\caption{Example of network with three stakeholders \(v_{A}\), \(v_{B}\) and \(v_{C}\), and connecting weighted edges. Adopted from~\cite{linaaker2016firms}.}
\label{fig:NetworkExample}
\end{figure}

If two stakeholder participated in an equal number of times, the size of each participation can be made further fine-grained. In another example, when considering an interaction network based on patches submitted to a software code repository, the size of a stakeholder's participation (\(X\)) can be quantified with the number of changed lines of code (LOC) of its patches. A simplified example is shown in Fig.~\ref{fig:NetworkExample} where three stakeholders \(v_A\), \(v_B\) and \(v_C\) each created various number of patches that were contributed to a certain issue. \(v_A\)'s patches contain 50 LOC in total. \(v_B\)'s patches contain 100 LOC in total, while \(v_C\)'s patches contain 150 LOC in total. Aggregated, 300 LOC were contributed to the issue. Resulting in the following edge weights: 

\begin{itemize}
    \item \(W(v_A,v_B) \& W(v_A,v_C)\ = 50/300\)
    \item \(W(v_B,v_A) \& W(v_B,v_C)\ = 100/300\) 
    \item\(W(v_C,v_A) \& W(v_C,v_B)\ = 150/300\)
\end{itemize}

By constructing this kind of networks (i.e., weighted and directed affiliation-networks~\cite{wasserman1994social,faust1997centrality}), stakeholders' interaction in an OSS ecosystem's RE process may be visualized on different abstraction levels across the different requirements artifact repositories identified in \textbf{S3}.

\textbf{Create influence profiles (S6):}
In a network, a stakeholder is more prominent if it has a central position with edges that make it extra visible and important to others~\cite{barnett2011encyclopedia}. In social networks, centrality measures are commonly used to analyze an actor's position and prominence relative to others~\cite{wasserman1994social}. Faust~\cite{faust1997centrality} breaks down the notion of centrality into how an actor is central given that they are active in the network, can communicate with others in the network efficiently, are able to mediate and control flows of information between others in the network, and have relationships with others that are central. These four aspects respectively relate to the centrality measures of out-degree, betweenness, closeness and eigenvector centrality. SIA uses these measures as the foundation for analyzing the influence of stakeholders.


These four centrality measures can be adapted in different ways to provide further facets of influence in regards to the interaction networks. As the interaction networks are described in \textbf{S5}, the edges that connect two stakeholders have weights attached to them. These weights allow the measures to take account of the relative size of each stakeholder's participation of the requirements artifacts on which the network is based on. E.g., out-degree centrality (see table~\ref{tbl:centralityMeasures}) refers to the sum of weights attached to outgoing edges from the focal stakeholder and its adjacent stakeholders~\cite{barrat2004architecture}. This gives an overall number in regards to the size of the focal stakeholder's participation in the set of requirements artifacts covered by the network. A high out-degree centrality may indicate that the focal stakeholder has a high influence on its adjacent neighbors and is good at communicating its views relative others in the network~\cite{orucevic2014network}. 
However, this way of measuring out-degree centrality does not provide information about the total number of connections of a stakeholder, which may better show the number of collaborations and opportunities to spread one's opinions~\cite{opsahl2010node}. Hence, we recommend that the proposed centrality measures are used both in the case where the edges have the relative weights attached to them, and in the case where they are considered either present or not~\cite{freeman1978centrality}. 


In table~\ref{tbl:centralityMeasures}, we describe the foundation for these measures and how they may be interpreted in terms of a stakeholder's influence in the RE process of an OSS ecosystem. 

As described by Faust~\cite{faust1997centrality}, centrality may be broken down into multiple aspects. Centrality measures, in turn, use different definitions and sets of criteria in regards to what classifies an actor's position as central. Hence, one measure can present a different social structure than another and different measures provide different perspectives on who are the most active~\cite{orucevic2014network}. In smaller and simpler network structures such measures may co-vary, while in larger and more complex networks, they may characterize actors very differently~\cite{hanneman2005introduction}.

Therefore, measures presented in table~\ref{tbl:centralityMeasures} could be seen as complementary to each other and may be used together to give each stakeholder (\(v_i\)) an \textit{influence profile} (\(IP_{v_i}\)), a 4-tuple consisting of each centrality measure (i.e., out-degree centrality (\(ODC_{v_i}\)), betweenness centrality (\(BC_{v_i}\)), closeness centrality (\(CC_{v_i}\)), eigenvector centrality (\(EC_{v_i}\))). 

\[IP_{v_i} = \left ( ODC_{v_i}, BC_{v_i}, CC_{v_i}, EC_{v_i} \right ) \]

Such a profile can then be used when analyzing a stakeholder's interaction network in step \textbf{S7}. E.g., a stakeholder in a certain interaction network may have

\begin{itemize}
    \item a high \(ODC\) indicating a high activity with many collaborations, 
    \item a low \(BC\) indicating that the stakeholder does not have a broker's position, but
    \item a high \(CC\) indicating that the stakeholder can more easily reach out with its communication, and 
    \item a high \(EC\) indicating that the stakeholder knows other influential stakeholders.
\end{itemize}

\begin{figure}[]
\centering
\includegraphics[scale=0.23]{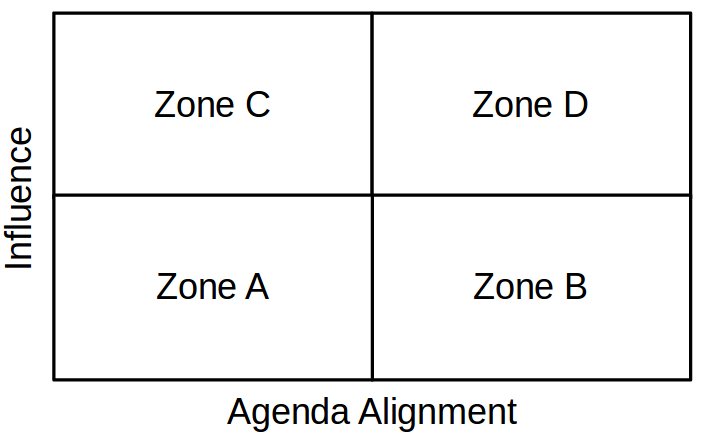}
\caption{Influence/Agenda alignment matrix to be used for stakeholder mapping. Adapted from earlier work~\cite{johnson2008exploring}.}
\label{fig:matrix}
\end{figure}

When comparing stakeholders and their influence profiles, it would be convenient to define, for each stakeholder \(v_i\), an aggregated \textit{influence score} \(IS_{v_{i}}\). Such a score could be used to divide stakeholders in to two groups, those with a high and low level of influence (see upper and lower zones in Fig. \ref{fig:matrix}). One way to do this aggregation is to simply add the normalized weights of each element in the profile, resulting in a ratio-scale number between $0$ and $1$, as given by the formula below, and then group stakeholders based on a threshold, eg. less than or equal to $0.5$ denotes low influence:
\begin{equation*}
IS_{v_{i}} = \frac{1}{4} \left ( \frac{ODC_{v_{i}}}{ODC_{max}} + \frac{BC_{v_{i}}}{BC_{max}} + \frac{CC_{v_{i}}}{CC_{max}} + \frac{EC_{v_{i}}}{EC_{max}} \right )
\end{equation*}

There are other ways of aggregating the different measures, using e.g. ordinal-scale ranks, a vector space distance metric (e.g. cosine similarity), a normalized exponential function (softmax), or applying some kind of weighting scheme to reflect e.g. that centrality is considered more interesting. Another option is to qualitatively compare the \(IS_{v_{i}}\) 4-tuple of measures in combination with some visualization technique, such as spider diagrams or similar. Future work should investigate which aggregation method that would best help to partition the stakeholders into high- and low-level category.    

In addition to comparing the stakeholders' influence profiles and overall influence scores within a specific stakeholder interaction network, it is equally important to compare between the networks. For example, if the analysis includes multiple requirements artifact repositories (e.g., issue-trackers and mailing-lists) or covers multiple releases, these could be cross-compared. A stakeholder may have a high overall influence score in one requirements artifact repository, and less in another. Further, the influence and interactions may shift with time why temporal analysis may give important insights. Also, it may be that one repository is more important than another (e.g., issue-tracker over mailing-list), as a result, the former should be given more attention in a cross-comparative analysis of a stakeholder. 


\textbf{Influence analysis of stakeholder interaction networks (S7):}
In the influence analysis, the interaction networks and influence profiles from \textbf{S5} and \textbf{S6} are used to address the purpose defined in \textbf{S1}. First, stakeholders are ranked on their overall influence score to get an overview of the stakeholder population. Stakeholders of interest, e.g., a top-list of those most influential, can then be cross-compared based on the centrality measures from their influence profiles, and analyzed in detail, e.g., in regards to their relationships. Table~\ref{tbl:centralityMeasures} provides descriptions of how the centrality measures may be interpreted in terms of a stakeholder's influence in the RE process of an OSS ecosystem. 

As a support in the analysis, and to help address the purpose as defined in \textbf{S1}, stakeholder mapping can be applied with the use of an influence/agenda alignment matrix (see Fig.~\ref{fig:matrix}). The matrix, based on earlier work~\cite{johnson2008exploring, mendelow1991stakeholder, newcombe2003client}, is adapted to consider the power and politics~\cite{frooman1999stakeholder} that play a central part in the RE process of OSS ecosystems~\cite{munir2017open, munir2015open}. The Y-axis represents the level of influence and the X-axis how well their agenda in the OSS ecosystem aligns with that of the focal firm. Both dimensions range from low to high. The four quadrants Zone A-D in the figure are explained subsequently.

The level of influence of a stakeholder is based on the influence score from \textbf{S6}. The threshold for when a stakeholder's influence score ranks as high is set by the analyst in relation to the total number of stakeholders in the network. Agenda alignment, which is the second dimension, is determined by qualitatively investigating the previously identified stakeholders' engagement in the OSS ecosystem, e.g., by reviewing comments made by the stakeholder in the set of issues which the analysis considers (as defined in \textbf{S1} and \textbf{S2}). The investigation should seek to answer if the stakeholder and the focal firm want the same thing, and to what extent.






The classification puts a stakeholder into one of four quadrants (A-D) of Fig.~\ref{fig:matrix}, each indicating a different relationship and possible engagement that the focal firm should establish and maintain with the stakeholder. Stakeholders with a high level of influence and high level of agenda alignment (Zone D) may pose as (potential) partners, both in regards to general collaboration and RE related activities and negotiations. Stakeholders with a high level of influence and low level of agenda alignment (Zone C) may pose as the key opponents and may require active engagement in negotiations in the RE process of the OSS ecosystem. Stakeholders with a low level of influence (Zone B and A) may not pose as having high importance, but may still require monitoring as they can move their position with time. Those in Zone B may pose as future collaboration opportunities, while those in Zone A as potential threats. 

If competitors are identified among those with high influence, this may signal that they have a high interest in the ecosystem and scope of the investigation. If they are found in Zone D, there might be an opportunity for co-opetition. In either case, whether they have aligning agendas or not, consideration should still be taken to the differential value of what is contributed and how resources are invested. By studying stakeholders in Zone C and D, a focal firm can potentially strengthen its own influence by learning from these stakeholders, in how they invest their resources and with whom they collaborate. This may lead to further collaboration and other potential partners, and how interest may overlap between multiple stakeholders.


\begin{table*}[htbp]
\centering
\caption{Network measures described from a general perspective as well as and how they can be interpreted from a RE perspective in regards to Stakeholder influence.}
\label{tbl:centralityMeasures}
\begin{tabular}[t]{p{1.5cm}p{5.5cm}p{9.5cm}}
\hline
Measure 
& Description 
& Stakeholder Influence Interpretation
\\ \hline
Out-degree Centrality
& Refers to how well-connected the focal actor is and considers the out-going edges towards its adjacent actors, where the focal actor is the transmitter (source) for the edges. With weights considered, this measure refers to the sum of weights attached to the outgoing edges of the focal actor~\cite{barrat2004architecture}. With binary edges considered, this measure refers to the number of outgoing edges between the focal actor and its adjacent actors~\cite{freeman1978centrality}. 

& Out-degree centrality is generally considered as a measure of activity that can identify "where the action is" and highlight the most visible actor in the network~\cite{wasserman1994social}. With weights considered, a high out-degree centrality is an indication of influence on adjacent stakeholders as the focal stakeholder has participated in a large part in the requirement artifacts which they have interacted with~\cite{orucevic2014network}. This participation can be viewed as the focal stakeholder's opinions in the RE process of the OSS ecosystem. In both cases of weighted and binary edges, a higher out-degree may also indicate a higher number of options or opportunities for qualitative contacts, i.e., to know the key stakeholders to influence and create traction with on a certain issue. For binary edges specifically, it may further indicate a high level of activity through a number of collaborations, but also to which the focal stakeholder has expressed its opinions.
\\\hline
Betweenness Centrality         
& Refers to the extent to which the focal actor lies on the shortest path between pairs of other actors. With weighted edges considered, it refers to the shortest path with the lowest sum of weights~\cite{brandes2001faster, newman2001scientific}. With binary edges considered, it refers to the shortest path in regards to the least number of edges~\cite{freeman1978centrality}.
& Betweenness centrality is a measure of control and coordination as it highlights actors who sit on the shortest, and sometimes only, communication paths or resource flow between many others~\cite{wasserman1994social}. Hence, stakeholders with a high betweenness centrality may control and coordinate the information flow about requirements, and interactions between other stakeholders. The focal stakeholder could be characterized as having a central position in the ecosystem, e.g., in regards to project management and governance. Others may be dependent on the focal stakeholders to relay the information and to set-up connections. Further, the centrality also indicates the ability to act as an intermediary that can influence the content of the information, and whom it reaches and when, to better serve personal priorities. When a stakeholder is the only one, or one of very few, linking two or more parts of a network, they are commonly referred to as brokers as their possibility to influence is very high~\cite{marczak2008information, newman2010networks}.
\\\hline
Closeness Centrality          
& Refers to the inverse of the sum of the shortest paths from the focal actor to all others in the network. With weighted edges considered, it refers to the shortest path with the lowest sum of weights~\cite{brandes2001faster, newman2001scientific}. With binary edges considered, it refers to the shortest path in regards to the least number of edges~\cite{freeman1978centrality}. This measure only considers those actors that are connected to the same network as the focal actor~\cite{newman2010networks}. For disconnected actors, the measure in undefined as the distance is infinite.  
& Closeness centrality is a measure of efficiency in contacting others and spreading, but also receiving, information in the network and hence an actors' ability to influence others~\cite{newman2010networks}. Hence, a high closeness centrality indicates that a stakeholder is efficient in spreading and receiving information about a requirement to and from the rest of the network of stakeholders. This efficiency allows the focal stakeholder to more easily communicate its agenda on the requirement and interact with others, e.g., in negotiations and lobbying. The focal stakeholder could, therefore, be characterized as being close to other stakeholders and more independent. This further minimizes the risk of intermediaries influencing the information about the requirement in an unfavorable manner~\cite{rowley1997moving}.
\\\hline
Eigenvector Centrality          
& Refers to how connected an actor is, similar to out-degree centrality, but considers how well-connected the adjacent actors are~\cite{bonacich1987power}. The focal actor receives a score based on a sum of its adjacent actors' scores~\cite{newman2010networks}.
& Eigenvector centrality is a measure of activity and visibility as out-degree centrality, but adds information to whom these attributes connect to. A high value indicates that the actor has important friends who in turn are visible and active~\cite{newman2010networks}. Hence, a high eigenvector centrality indicates that a stakeholder knows and collaborates with other stakeholders who are important and have key positions in the OSS ecosystem~\cite{faust1997centrality}. The focal stakeholder is in a position to have a potentially high impact on the RE process in the ecosystem by being able to communicate its agenda to, and influence key actors in the social network~\cite{faust1997centrality}.
\\\hline
\end{tabular}
\end{table*}

\section{Case Study of Apache Hadoop OSS Ecosystem}
\label{sec:casestudy}

In this section, we describe a first evaluation of SIA in our design methodology. We demonstrate the applicability and utility of SIA in a case study ~\cite{runeson2012casestudy} on the Apache Hadoop OSS ecosystem. The case study takes the perspective of a (fictive) focal firm that provides scalable and secure infrastructure on which Hadoop can be deployed for customers. This is a new product offering, and the focal firm is now interested in becoming active in the Apache Hadoop OSS ecosystem. As they are new to the ecosystem, they want to do an initial stakeholder analysis to see if there are any potential partners to collaborate with, and potentially learn from (\textbf{S1}). First, they want to get a general overview of the stakeholder population to see who is present and how the ecosystem functions in terms of the power structure and collaboration patterns. Second, they will look for potential partners among those most influential and investigate how they work, and what interests they have in the ecosystem.

The Apache Hadoop project\footnote{http://hadoop.apache.org/} is a widely adopted OSS framework for distribution and process parallelization of large data, originating from \textit{Yahoo} in 2006. The framework consists of four modules:
Hadoop Common Modules, Hadoop Distributed File System (HDFS), Hadoop YARN, and Hadoop MapReduce.


\begin{figure}[]
    \centering
    \includegraphics[width=\columnwidth]{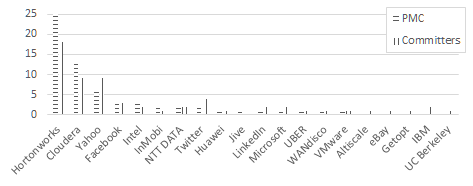}
    \caption{Number of committers and members in the Apache Hadoop PMC aggregated per firm.}
    \label{fig:governanceDiagram}
\end{figure}

The Apache Hadoop project is part of the Apache Software Foundation which is an umbrella organization for a large number of OSS projects and their ecosystems. A common trait for these projects is the use of meritocracy in terms of culture and governance\footnote{https://www.apache.org/foundation/how-it-works.html\#meritocracy}. This is reflected in the governance structure among the Apache projects, as in Apache Hadoop which is governed by a Program Management Committee (PMC) that consists of representatives from the Apache Software Foundation and of elected members from the project's ecosystem. Further, the PMC members are also classified as committers, i.e., they have been granted write access to the project. A member may be elected as a new committer by the existing ones. Being elected as a committer does however not imply membership of the PMC. To become a committer or member of the PMC, an individual need to show merit, e.g., by contributing and actively participating in the development of the project. Hence, power may be earned by showing a long-term commitment and having the competence needed (i.e., meritocracy). In Fig.~\ref{fig:governanceDiagram} the distribution of members of the committers and the PMC are presented based on affiliation per firm.



\subsection{Overview of Stakeholder Interaction and Influence}
To get a recent view on who the most influential stakeholders are, the scope of the analysis is limited to requirements included from release 2.2.0 (15/Oct/13) to 2.7.1 (06/Jul/15) (\textbf{S2}). To get a view on both social and technical interaction, the issue-tracker is analyzed in regards to requirements artifact repositories (\textbf{S3}). The issues contain both comments (the social dimension) and patches (technical). The patches are committed by authorized users, once they have been approved. To identify the organizational affiliation of individuals that have interacted via the requirements (\textbf{S4}), an analysis is done of e-mail subdomains, complemented with cross-checking against other information sources (e.g., social media and electronic archives)~\cite{bird2012examining,gonzalez2013understanding, linaaker2016firms}. For a subset of individuals, an organizational affiliation could not be determined. These individuals were aggregated into two separate groups, either independent (if this could be determined) or unidentified.

\textbf{Creation of Stakeholder Interaction Networks (S5): }
Based on the scope specified in \textbf{S2}, and the repository identified in \textbf{S3}, two interaction networks are generated: a \textit{comments-network} to include stakeholders who commented on common issues, and a \textit{patch-network} to include the stakeholders who contributed patches to the same issues (\textbf{S5}). The patch-network was presented in earlier work~\cite{linaaker2016firms}, and a similar data collection and cleaning approach were used in order to create the comments-network, as is also proposed in SIA (see section~\ref{sec:SIA}). The comments-network shows activity and collaboration of a stakeholder in regards to the social interaction and discussion that revolves around a certain issue, and the patch-network shows same characteristics for a stakeholder in regards to suggesting technical implementations. 

\begin{figure*}
\centering
\begin{subfigure}{.50\textwidth}
    \includegraphics[width=\columnwidth]{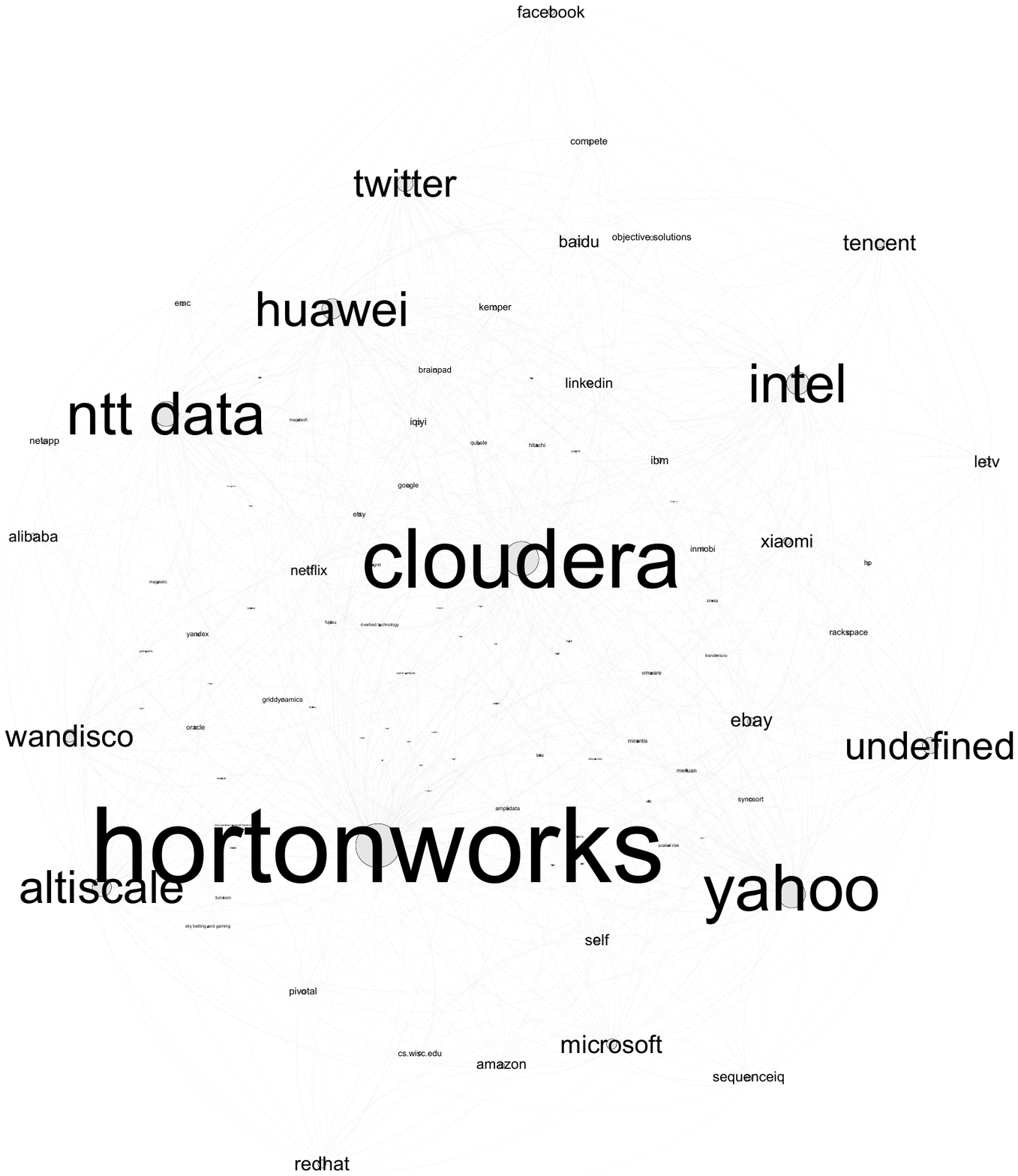}
    \caption{Comments-network.}
    \label{fig:commentsNetwork}
\end{subfigure}%
\begin{subfigure}{.5\textwidth}
    \includegraphics[width=\columnwidth]{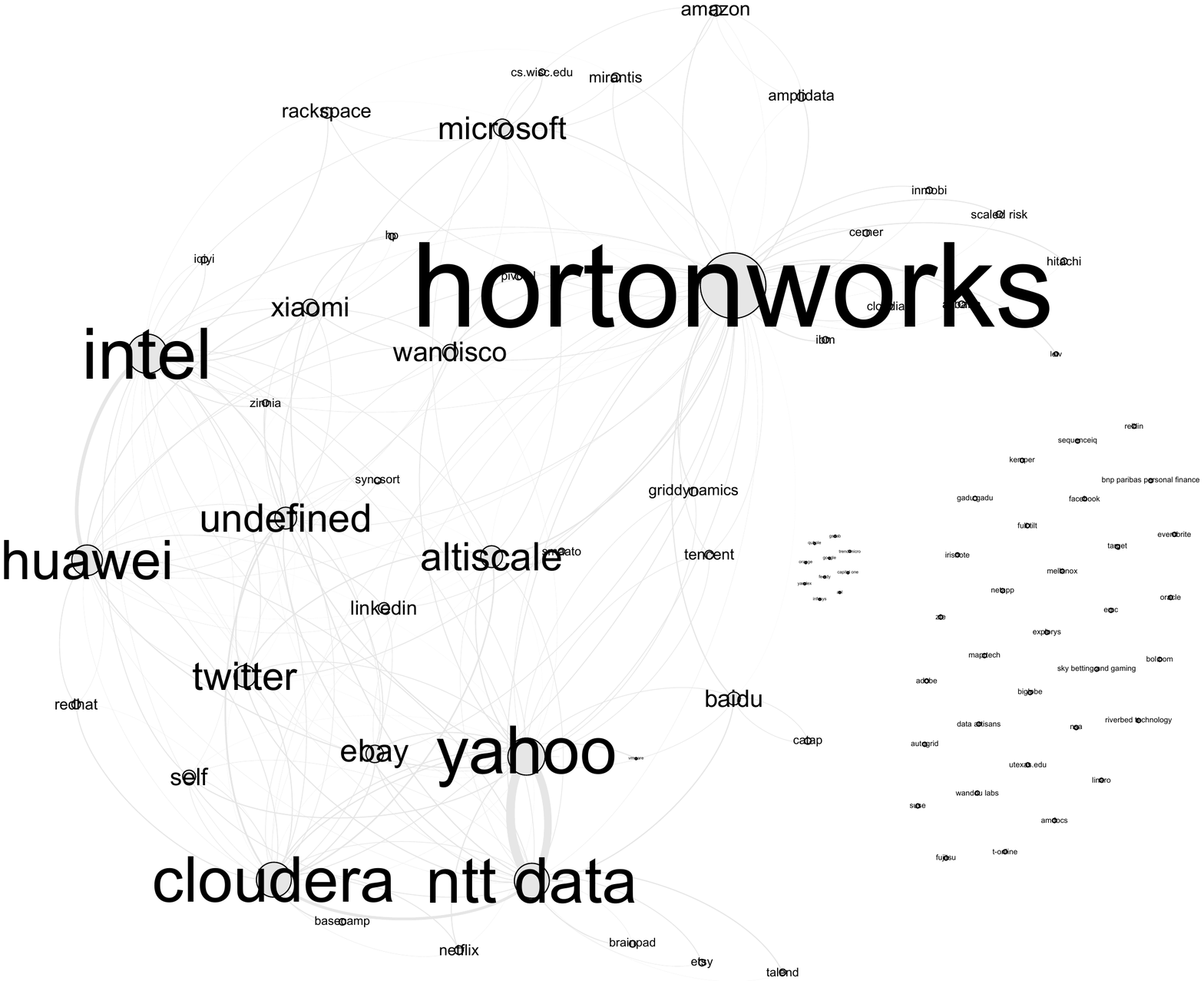}
    \caption{Patch-network.}
    \label{fig:patchNetwork}
\end{subfigure}
\caption{Visualization of the (a) comments- and (b) patch-networks. Labels are of firms and of relative size of their weighted out-degree to other firms in each network.}
\label{fig:networkGraphs}
\end{figure*}

In each of the two networks, a stakeholder is represented by a node, and the collaborations between them are represented by the edges connecting the nodes. The comments-network consists of 122 stakeholders, compared to 86 stakeholders in the patch-network (see table~\ref{tbl:networkData}). In both cases, this includes two groups of developers classified as independent or as unidentified. The comments-network has a higher degree of collaboration with an average of 9 collaborations per stakeholder, compared to the patch-network, which has an average of 3 collaborations per stakeholder. Both networks are visualized on a high level in Fig.~\ref{fig:commentsNetwork} and~\ref{fig:patchNetwork}. Labels are of firms and of relative size to their weighted out-degree, a reason for which only those with the highest values may be readable.

\begin{table}[htbp]
\centering
\caption{Characteristics of comments- and patch-networks.}
\label{tbl:networkData}
\begin{tabular}{@{}lll@{}}
\toprule
& Comments-network & Patch-network 
\\ 
\midrule
Stakeholders & 122 & 86
\\
Collaborations & 1096 & 260 
\\
Per stakeholder & 9 & 3              
\\ \bottomrule
\end{tabular}
\end{table}

\begin{figure*}
\centering
\begin{subfigure}{.5\textwidth}
    \includegraphics[width=\columnwidth]{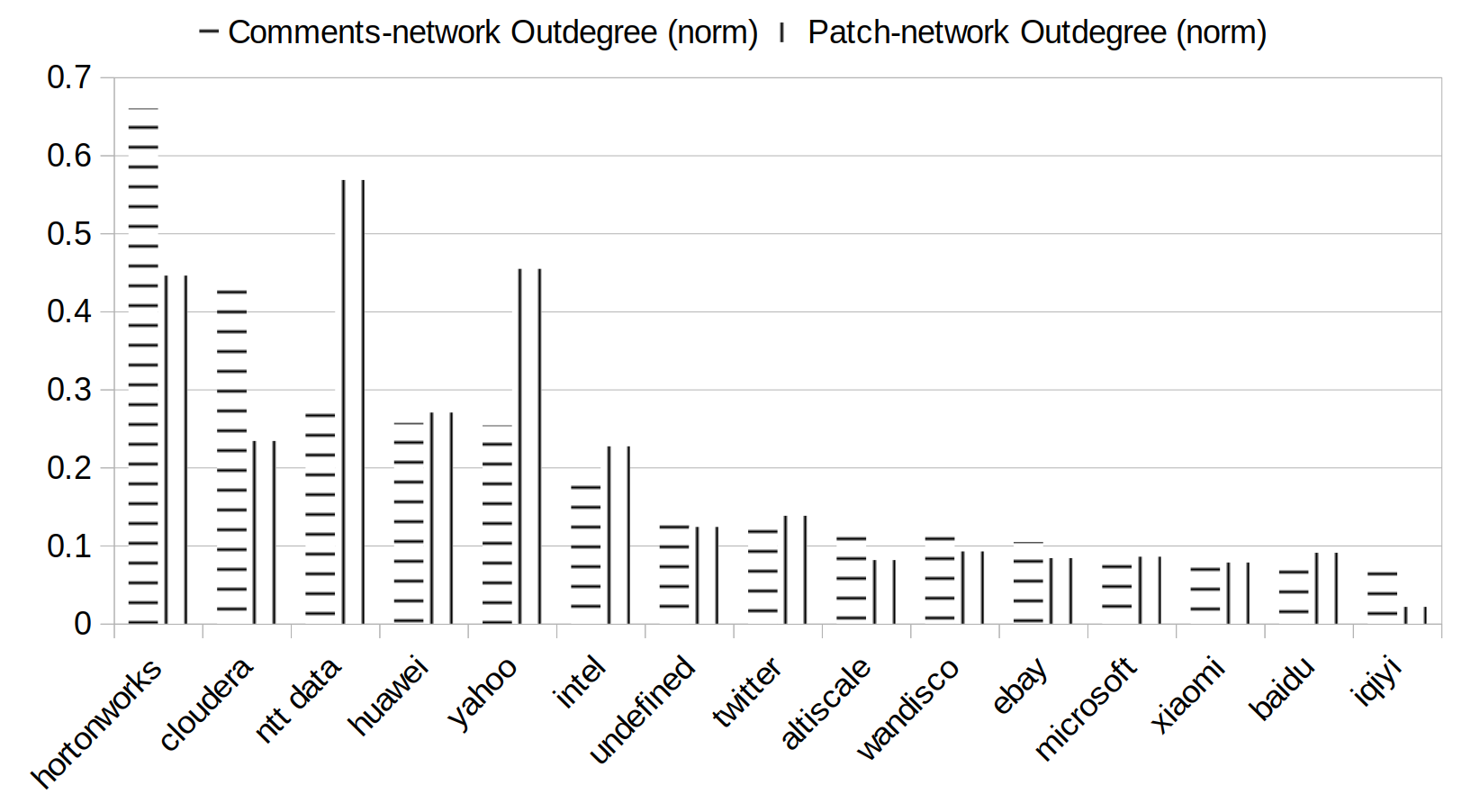}
    \caption{Out-degree centrality.}
    \label{fig:outdegreeDiagram}
\end{subfigure}%
\begin{subfigure}{.5\textwidth}
    \includegraphics[width=\columnwidth]{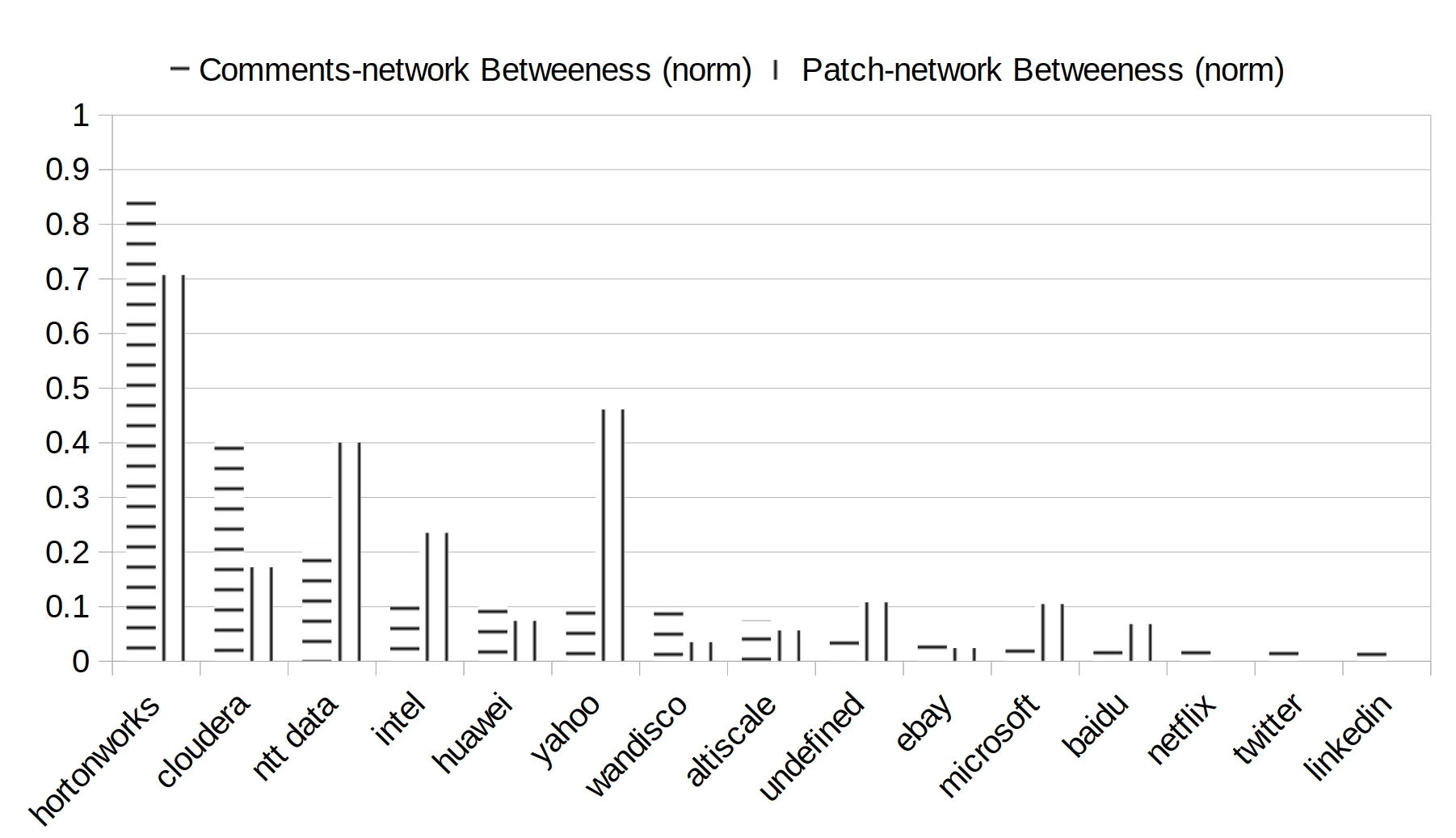}
    \caption{Betweeness centrality.}
    \label{fig:betweenessDiagram}
\end{subfigure}
\caption{Visualizations of normalized centrality measures for the 15 top influential firms across the comments- and patch-networks. Each diagram is sorted in a descending order based on respective centrality measure from the comments-network.}
\label{fig:centralityDiagrams}
\end{figure*}

\textbf{Creation of influence profiles. (S6): }To measure the influence of, and collaboration among, the stakeholders (\textbf{S6}), two SNA measures were leveraged: weighted out-degree and betweenness centrality. Other centrality measures presented in table~\ref{tbl:centralityMeasures} were excluded due to space considerations in this paper. In Fig.~\ref{fig:centralityDiagrams}, the two measures are presented in two separate diagrams. The diagrams contrast the respective measures for the comments- and patch-networks in regards to the 15 top stakeholders (considering the overall influence score). 

\textbf{Influence Analysis of the Stakeholder Interaction Networks (S7): }
As presented in table~\ref{tbl:centralityMeasures}, the measures measure different aspects of influence and collaboration among the stakeholders. Below, the two measures are compared in regards to the two networks and their stakeholders.

\textit{Out-degree centrality: }Fig.~\ref{fig:outdegreeDiagram} illustrates the normalized out-degree centrality which may be considered as rather equal for most stakeholders with the exception of those most influential: NTT Data, Yahoo, Hortonworks, and Cloudera. Both NTT Data and Yahoo have a notably higher influence in regards to technical implementation-suggestions, while Hortonworks and Cloudera have a higher influence and activity through social interaction and discussion. Considering the distribution of stakeholders from the different user categories, a heavier representation of product vendors (Hortonworks, Cloudera, and Huawei) can be seen in the top five, in regards to both the comments- and patch-networks.

\textit{Betweeness centrality: }In Fig.~\ref{fig:betweenessDiagram}, it can be seen that the normalized betweenness centrality varies notably between the comments- and patch-networks for the top stakeholders. Hortonworks has the highest betweenness centrality in regards to both the technical and social aspects and compared to Cloudera and Yahoo, it has double the betweenness centrality in the comments- and patch-networks respectively. Contrasting Cloudera and Yahoo, a clear difference in focus and importance is shown. Cloudera values technical implementation suggestions over social interaction and discussion, while Yahoo focuses on social interaction and discussions. 


\begin{table}[htbp]
\centering
\caption{Top ten stakeholders based on influence score based on comment-network, considering only the out-degree and betweenness centrality.}
\label{tbl:influenceScore}
\begin{tabular}{p{1.5cm}p{1.2cm}p{1.2cm}p{1.2cm}p{1.2cm}}
\toprule
Stakeholder & Outdegree (norm) & Betweeness (norm)  & Influence score \\ \midrule
hortonworks & 0.66 & 0.86 & 0.76 \\
cloudera & 0.44 & 0.4 & 0.42 \\
ntt data & 0.28 & 0.22 & 0.25 \\
huawei & 0.26 & .10 & 0.18 \\
yahoo & 0.25 & 0.10 & 0.18 \\
intel & 0.19 & 0.11 & 0.15 \\
undefined & 0.11 & 0.09 & 0.10 \\
twitter & 0.14 & 0.06 & 0.10 \\
altiscale & 0.11 & 0.07 & 0.09 \\
wandisco & 0.13 & 0.02 & 0.8 \\ \bottomrule
\end{tabular}
\end{table}



\textit{Cross-comparison of centrality measures: }
To simplify the cross-comparison, the influence score is used to get an overview of the top 10 most influential stakeholders considering the two centrality measures (see table~\ref{tbl:influenceScore}). Comparing the two centrality measures for these ten, both similarities and differences may be found. Although it has higher activity in the comments-network, Hortonworks has high influence in regards to both technical and social interaction, if both centrality measures are taken into account. This indicates that Hortonworks has a high impact in regards to what is implemented and how. This firm can be classified as well-connected both directly and indirectly and has a good position to act as an authority in regards to information spread and coordination. NTT Data and Yahoo both clearly have a higher degree of activity and influence in the patch-network. As with Hortonworks, they also have a similar distribution among both of the two measures. This may indicate that they have a high impact in regards to what is implemented and how, but focus their resources on contributing technical implementation suggestions and solutions. As with Hortonworks, they can be classified as well connected both directly and indirectly, and have a good position to act as an authority in regards to information spread and coordination. Regarding the out-degree centrality, a group of stakeholders forms just below the top. 

Considering the influence/agenda alignment matrix (see Fig.~\ref{fig:matrix}), these stakeholders could be considered as key stakeholders and qualify for either Zone C or D. They could pose either as potential partners or threats depending on how their agenda aligns. Also, depending on if they are competitors or not, consideration should also be taken when constructing contribution strategies~\cite{wnuk2012can}. The focal firm should, therefore, monitor and form an understanding of how these stakeholders' agendas align with their own.


\subsection{Investigating Collaborations and Agenda of a Potential Partner}
\label{sec:Case:DeeperInv}
From the previous analysis, the focal firm could identify WANdisco as a stakeholder with a similar business model and a potential partner in terms of collaboration and similar interests (Zone D in Fig.~\ref{fig:matrix}). The goal in this second step is to do a more thorough analysis focusing on WANdisco's collaborations and high-level agenda (\textbf{S7}).

Looking at WANdisco's influence profile, their overall influence score gives them an ordinal rank of 10 (see table~\ref{tbl:influenceScore}) when analyzing the comments network. They have an equal level of social and technical activity, on similar levels as Twitter, Altiscale, eBay and Microsoft (see Fig.~\ref{fig:outdegreeDiagram}). They have a relatively high level of control and coordination considering the betweenness centrality. 
All things considered, they have a relatively high influence and interest in Apache Hadoop, but much lower than the key-stone players, Hortonworks, Cloudera, NTT Data, Huawei, Yahoo, and Intel.

WANdisco entered the Apache Hadoop ecosystem in 2012 by acquiring AltoStar. Their product is a platform that allows for distribution of data over multiple Apache Hadoop clusters, similar to that of the focal firm. WANdisco has 14 active developers in the investigated set of releases in regards to comments and patch-contributions. One developer is also a member of the PMC and Committers group. 

To learn more about WANdisco's interests in Apache Hadoop and its collaborators, the focal firm investigates if WANdisco has shown a special focus in regards to any of the four modules of Apache Hadoop: Common, HDFS, YARN, and MapReduce (\textbf{S2}). The analysis is still focused on requirements included in releases R2.2-R2.7. In regards \textbf{S3-4}, they are identical to the previous example. When creating the interaction networks (\textbf{S5}), one patch- and one comments-network are created for each of the modules.

When creating the influence profiles (\textbf{S6}), the analysis is limited to examining the out-degree to get a view of their activity and comprehension of their relative influence in regards to the modules. Values for binary and weighted out-degree are presented in  table~\ref{tbl:binOutdegreeWandisco} and table~\ref{tbl:weightedOutdegreeWandisco} respectively. The former specifically indicates the number of other stakeholders that WANdisco has interacted with, and the latter a better relative measure of their influence. As can be noticed for values regarding the patch-network, it can be concluded that WANdisco has a specific interest in the HDFS module. The out-degree values for the comments network further confirm a specific interest in the HDFS module with a relative ranking of 5 and 6 respectively out of 48. Some interest can also be observed for the Common module.

\begin{table}[htbp]
\centering
\caption{Binary out-degree of Wandisco for Apache Hadoops four modules. Values aggregated for releases R2.2-2.7 per network type. Relative ranking within parenthesis.}
\label{tbl:binOutdegreeWandisco}
\begin{tabular}{p{1cm}p{1.3cm}p{1.2cm}p{1.2cm}p{1.2cm}}
\toprule
 & Common & HDFS & YARN & Mapreduce \\ \midrule
Comments & 11 (11/64) & 20 (5/48) & 4 (32/59) & 1 (33/39) \\
Patches & 0 & 5 (7/24) & 0 & 0 \\ \bottomrule
\end{tabular}
\end{table}

\begin{table}[htbp]
\centering
\caption{Weighted out-degree of Wandisco for Apache Hadoops four modules. Values aggregated for releases R2.2-2.7 per network type. Relative ranking within parenthesis.}
\label{tbl:weightedOutdegreeWandisco}
\begin{tabular}{p{1cm}p{1.3cm}p{1.2cm}p{1.2cm}p{1.2cm}}
\toprule
 & Common & HDFS & YARN & Mapreduce \\ \midrule
Comments & 1.87 (12/64) & 2,82 (6/48) & 0.73 (19/59) & 0.24 (26/39) \\
Patches & 0 & 2.97 (7/24) & 0 & 0 \\ \bottomrule
\end{tabular}
\end{table}

Regarding collaboration, the analysis is limited to the HDFS module as this is where their main interest of WANdisco lies. In regards to the patch-network, there are only five collaborators, as indicated by the binary out-degree in table~\ref{tbl:binOutdegreeWandisco}. These consist of Hortonworks, Huawei, Intel, Yahoo, and Intel. In regards to comments-network, WANdisco had interacted with 20 other stakeholders. Out of these, Hortonworks, Cloudera, Intel, Pivotal and Yahoo were the top five in regards to the number of comments made by WANdisco on common issues, see table~\ref{tbl:commentsCollaboratorsWandisco}. The table further presents the weight of the outgoing edge from WANdisco to each respective stakeholder. In the example of Pivotal, this may be interpreted as WANdisco having made 53 percent of the total number of comments on issues where both WANdisco and Pivotal have collaborated on.

An outcome of this analysis is that WANdisco holds their main interest and invest their resources in the HDFS component, both from a technical and social perspective. Considering the influence/agenda alignment matrix in \textbf{S7} (see Fig.~\ref{fig:matrix}), a qualitative investigation needs to be performed, e.g., of their comments and code commits, in order to determine e.g., what features they value or prioritize. Such an investigation will help to determine the agenda alignment further and if WANdisco belongs to Zone C or D, i.e., if they make up a potential opponent or partner. Based on their active collaborations, Pivotal should be investigated further in terms of their interest and activity.

\begin{table}[htbp]
\centering
\caption{Top collaborators with Wandisco in the comments network of the HDFS module.}
\label{tbl:commentsCollaboratorsWandisco}
\begin{tabular}{p{1.5cm}p{1.5cm}p{2cm}p{1.5cm}}
\toprule
Stakeholder & Number of comments & Total number of comments & Weight \\ \midrule
Hortonworks & 227 & 1109 & 0.20 \\
Cloudera & 98 & 663 & 0.15 \\
Intel & 91 & 679 & 0.13 \\
Pivotal & 42 & 79 & 0.53 \\
Yahoo & 34 & 313 & 0.11 \\ \bottomrule
\end{tabular}
\end{table}

\section{Discussion}
\label{sec:discussion}

Below we first discuss different alternatives to characterizing a stakeholder's influence, followed by a discussion regarding limitations and threats to validity in our demonstration of SIA's utility in the analysis of the Apache Hadoop ecosystem stakeholder influence.

\subsection{Alternatives to Characterizing a Stakeholder's Influence}
\label{sec:Disc:Usefullness}
The three questions stipulated by Frooman~\cite{frooman1999stakeholder} highlights the strategic importance of stakeholder identification and analysis: firms need to identify and characterize present stakeholders in terms of their influence, identify their agendas primarily in terms of alignment with one's own, and how they are planning to achieve it. The latter of the three is important as it informs of a firm's possible strategies for contribution and interaction. SIA helps to address these questions by characterizing the stakeholders' collaboration and influence within the OSS ecosystem. The quantitative outcome which is generated is best complemented with qualitative insights which may be gained through observing or even taking part in the communication of the ecosystem.

In the stakeholder mapping process, which is part of the influence analysis of SIA (\textbf{S7}), both the quantitative and qualitative aspects are needed. In the proposed influence/agenda alignment matrix, the influence profiles and influence scores may be used to determine the level of influence. As mentioned i Section~\ref{sec:SIA} and the description of (\textbf{S6}), the proposed influence score is one approach to get a simplified overview of which stakeholders are the most influential. However, as the different centrality measures provide different aspects, these measures should still be investigated qualitatively to get a more fair view over how influential the stakeholders can be considered to be relative others.

On a general level, one can also look at which stakeholders hold a seat in the different committees of the ecosystem governance structure. However, these do not have to align as a firm can influence both by having representatives in and outside leadership positions, of which the latter is the more common~\cite{schaarschmidt2015firms}. This phenomenon can be noted when comparing firms with members on the PMC and Committers group in the Apache Hadoop OSS ecosystem (see Fig.~\ref{fig:governanceDiagram}) with firms that have the overall highest influence score (based on outdegree and betweenness centrality, see table~\ref{tbl:influenceScore}). NTT Data, with a relatively high activity both in regards to the patch and comments networks (see Fig.~\ref{fig:networkGraphs}), have a very limited number of places on both the PMC and Committers group. Furthermore, when changing the scope of the analysis (e.g., a certain set of releases or a component - see Section~\ref{sec:Case:DeeperInv}), the governance structure may give an even less representative overview as different stakeholders have different interests, why the network approach proposed by SIA may prove more valuable. 

Another approach to measuring influence with other than centrality measures would be to use pure count-based measures of a developer's activity, e.g., number of comments and code-commits. As highlighted by Joblin et al.~\cite{joblin2017classifying}, these, however, give a simplified view of a developers position and do not consider the inter-developer relationship. By considering the latter, an analysis can investigate, for example, how active the developers are, how efficiently they can communicate with others, how they are able to mediate and control the flow of information between others, and have relationships ''with others that are themselves'' central~\cite{faust1997centrality}. As further shown~\cite{joblin2017classifying}, network-based measures are equally good, and in certain cases better than count-based measures at describing how influential a developer is. Certain count-based aspects are however included in SIA as it does recommend the use of binary edges as a complement to the weighted edges. As is mentioned in S6 (see Section~\ref{sec:SIA}), a high out-degree centrality based on weighted edges may indicate that the focal stakeholder has a high influence on its adjacent neighbors and is good at communicating its views relative others in the network~\cite{orucevic2014network}. However, this way of measuring out-degree centrality does not provide information about the total number of connections of a stakeholder, which may better show the number of collaborations and opportunities to spread one's opinions~\cite{opsahl2010node}.

Furthermore, it may be noted that there are other centrality measures available~\cite{wasserman1994social} than those proposed in the CSF. We based our choice of out-degree, betweenness, closeness and eigenvector centrality measures on the suggestion of Faust~\cite{faust1997centrality} as explained in section~\ref{sec:SIA}. These are generally adopted in explaining the centrality and importance of an actor when analyzing OSS ecosystems (e.g., ~\cite{bird2006mining, hossain2006actor, orucevic2014network}).

\subsection{Limitations and Threats to Validity}

As a proof of concept demonstration that SIA is functional and practical in stakeholder analysis in a large ecosystem, we described a case study on the Apache Hadoop OSS ecosystem. The ecosystem has a community-managed governance model, meaning that the OSS project is owned and managed by the community~\cite{o2007governance}, and a meritocratic authority structure, meaning that influence is gained by proving merit~\cite{nakakoji2002evolution} and by establishing a symbiotic relationship with the ecosystem~\cite{dahlander2005relationships}. Another important characteristic of the chosen ecosystem is the high concentration of firms among its stakeholders, as we are interested in identifying and analyzing stakeholder on the organizational level.


This application of SIA in our case study, however, is not without a number of threats to validity. A threat to the internal validity concerns the way how weights are calculated (see S5, Section~\ref{sec:SIA}). The consideration taken to the relative size in regards to changed lines of code does account for the net amount (i.e., added and removed lines), but commits containing larger amounts of non-meaningful content may give a non-fair view. Thus it may be valuable to compare interaction networks and influence profiles based on both weighted and binary edges. Also, comparing networks based on different requirement artifact repositories (as exemplified in Section~\ref{sec:casestudy}) can help to give a more nuanced view.

A related threat is that we consider issues in general as "requirements", which may be further extended in our reasoning of requirements artifacts in general. This is based on the nature of RE in OSS as informal and decentralized~\cite{ernst2012case}. Requirements consist of fragmented representations, such as issues, mail-thread discussions and commits~\cite{scacchi2002understanding}. Further mitigation of this threat could include textual and natural language processing of the content in each of the requirements artifacts. This is a vibrant topic in the research field of mining software repositories. However, we consider this topic as out of the scope of SIA as we focus on the identification and stakeholder analysis process in its form and structure. We do acknowledge the topic as complementary quality aspects that should be further researched and integrated with our proposed process in future research.

A further threat concerns the determination of organizational affiliation of individuals in the OSS ecosystem. We adopted a heuristic approach as suggested by earlier research~\cite{bird2012examining,gonzalez2013understanding}, starting with an analysis of email sub-domains and complementing with second and third level sources such as social network sites as LinkedIn and Facebook, as well as blogs, community communication (e.g., comment-history, mailing-lists, IRC logs), web articles and firm websites. We acknowledge this is a delicate and complex process that is best mitigated by "knowing" the ecosystem and actively interacting with its communication channels. In SIA, we recommend using a mix-method triangulation with both qualitative and quantitative approaches.




The case study we described exemplifies how the different steps of SIA can be applied and the insights that can be gained. We acknowledge that a single case study is not sufficient to prove validity in terms of repeatability and utility, and that this is only a first step in the artifact validation phase of our design science research~\cite{wieringa2014design}. The characteristics of the Apache Hadoop OSS ecosystem, i.e. community-managed, meritocratic and multi-vendor, do however indicate what types of OSS ecosystems might benefit from a stakeholder analysis using SIA. Further investigation of SIA's utility and repeatability is out of the scope of this study and instead left for future work. Future research should consider applying SIA from a focal firm's perspective and study different types of OSS ecosystems with a more nuanced authority structure, e.g., as both autocratic, democratic and meritocratic coordination processes can act in parallel~\cite{shaikh2017governing}. This falls naturally in the design science research approach as it is an iterative search process for an artifact that will solve the stated problem~\cite{hevner2004design}.



\section{Conclusions}
\label{sec:conclusions}

This study proposes the Stakeholder Influence Analysis (SIA) method which aims to help firms involved in OSS ecosystems to characterize ecosystem's stakeholders according to their level of influence on the ecosystem's RE process. This is an important attribute due to the collaborative and informal nature of the OSS ecosystem's RE processes, and often meritocratic governance structure. SIA, therefore, allows firms to see in which requirements a stakeholder holds a certain interest, and thereby create an overview of a stakeholder's agenda. This also allows firms to understand how stakeholders invest their resources, and with whom they collaborate according to their agenda. Thus, SIA offers input to how firms involved in OSS ecosystems should construct their contribution strategies and act in the politics and negotiations of the ecosystem's RE process in order to align it with their internal RE process and product planning. It can be concluded that SIA shows potential through the case study on the Apache Hadoop OSS ecosystem, while further work is needed in regards to external validity.


In future work, we therefore aim to refine and validate SIA quantitatively and qualitatively through further design cycles involving additional OSS ecosystems, and from existing focal firms' perspectives. Further, we aim to investigate how the stakeholder analysis processes resulting from SIA may be used as an input to the construction and execution of contribution strategies~\cite{wnuk2012can}.

\textbf{Acknowledgements:} We would like to thank Dr. Alma Orucevic-Alagic and the anonymous reviewers for their valuable feedback and inputs.

\bibliographystyle{unsrt}
\bibliography{bibliography_master}

\begin{thebibliography}{10}

\bibitem{munir2015open}
Hussan Munir, Krzysztof Wnuk, and Per Runeson.
\newblock Open innovation in software engineering: a systematic mapping study.
\newblock {\em Empirical Software Engineering}, 21(2):684--723, Apr 2016.

\bibitem{nakakoji2002evolution}
Kumiyo Nakakoji, Yasuhiro Yamamoto, Yoshiyuki Nishinaka, Kouichi Kishida, and
  Yunwen Ye.
\newblock Evolution patterns of open-source software systems and communities.
\newblock In {\em Proceedings of the international workshop on Principles of
  software evolution}, pages 76--85. ACM, 2002.

\bibitem{jansen2009business}
Slinger Jansen, Sjaak Brinkkemper, and Anthony Finkelstein.
\newblock Business network management as a survival strategy: A tale of two
  software ecosystems.
\newblock {\em Proccedings of the 1st International Workshop on Software
  Ecosystems}, pages 34--48, 2009.

\bibitem{glinz2007guest}
Martin Glinz and Roel~J Wieringa.
\newblock Guest editors' introduction: Stakeholders in requirements
  engineering.
\newblock {\em IEEE Software}, 24(2):18--20, 2007.

\bibitem{alspaugh2013ongoing}
Thomas Alspaugh, Walt Scacchi, et~al.
\newblock Ongoing software development without classical requirements.
\newblock In {\em 21st IEEE International Requirements Engineering Conference},
  pages 165--174. IEEE, 2013.

\bibitem{ernst2012case}
Neil Ernst and Gail~C Murphy.
\newblock Case studies in just-in-time requirements analysis.
\newblock In {\em IEEE Second International Workshop on Empirical Requirements
  Engineering}, pages 25--32. IEEE, 2012.

\bibitem{scacchi2002understanding}
Walt Scacchi.
\newblock Understanding the requirements for developing open source software
  systems.
\newblock In {\em Software, IEE Proceedings-}, volume 149, pages 24--39. IET,
  2002.

\bibitem{german2003gnome}
Daniel~M German.
\newblock The gnome project: a case study of open source, global software
  development.
\newblock {\em Software Process: Improvement and Practice}, 8(4):201--215,
  2003.

\bibitem{laurent2009lessons}
Paula Laurent and Jane Cleland-Huang.
\newblock Lessons learned from open source projects for facilitating online
  requirements processes.
\newblock In Martin Glinz and Patrick Heymans, editors, {\em Requirements
  Engineering: Foundation for Software Quality}, pages 240--255, Berlin,
  Heidelberg, 2009. Springer Berlin Heidelberg.

\bibitem{baars2012framework}
Alfred Baars and Slinger Jansen.
\newblock A framework for software ecosystem governance.
\newblock In Michael~A. Cusumano, Bala Iyer, and N.~Venkatraman, editors, {\em
  Software Business}, pages 168--180, Berlin, Heidelberg, 2012. Springer Berlin
  Heidelberg.

\bibitem{dahlander2005relationships}
Linus Dahlander and Mats~G. Magnusson.
\newblock Relationships between open source software companies and communities:
  Observations from nordic firms.
\newblock {\em Research Policy}, 34(4):481 -- 493, 2005.

\bibitem{jensen2007role}
Chris Jensen and Walt Scacchi.
\newblock Role migration and advancement processes in ossd projects: A
  comparative case study.
\newblock In {\em 29th International Conference on Software Engineering, 2007},
  pages 364--374. IEEE, 2007.

\bibitem{wnuk2012can}
Krzysztof Wnuk, Dietmar Pfahl, David Callele, and Even-Andr{\'e} Karlsson.
\newblock How can open source software development help requirements management
  gain the potential of open innovation: an exploratory study.
\newblock In {\em Proceedings of the ACM-IEEE international symposium on
  Empirical software engineering and measurement}, pages 271--280. ACM, 2012.

\bibitem{frooman1999stakeholder}
Jeff Frooman.
\newblock Stakeholder influence strategies.
\newblock {\em Academy of management review}, 24(2):191--205, 1999.

\bibitem{rowley1997moving}
Timothy~J Rowley.
\newblock Moving beyond dyadic ties: A network theory of stakeholder
  influences.
\newblock {\em Academy of management Review}, 22(4):887--910, 1997.

\bibitem{Milne2012}
Alastair Milne and Neil Maiden.
\newblock Power and politics in requirements engineering: embracing the dark
  side?
\newblock {\em Requirements Engineering}, 17(2):83--98, 2012.

\bibitem{aurum2003fundamental}
Ayb{\"u}ke Aurum and Claes Wohlin.
\newblock The fundamental nature of requirements engineering activities as a
  decision-making process.
\newblock {\em Information and Software Technology}, 45(14):945--954, 2003.

\bibitem{pacheco2012systematic}
Carla Pacheco and Ivan Garcia.
\newblock A systematic literature review of stakeholder identification methods
  in requirements elicitation.
\newblock {\em Journal of Systems and Software}, 85(9):2171--2181, 2012.

\bibitem{freeman2010strategic}
R~Edward Freeman.
\newblock {\em Strategic management: A stakeholder approach}.
\newblock Cambridge University Press, 1984.

\bibitem{wieringa2014design}
Roel~J Wieringa.
\newblock {\em Design science methodology for information systems and software
  engineering}.
\newblock Springer, 2014.

\bibitem{hevner2004design}
Alan~R Hevner, Salvatore~T March, Jinsoo Park, and Sudha Ram.
\newblock Design science in information systems research.
\newblock {\em MIS quarterly}, 28(1):75--105, 2004.

\bibitem{wasserman1994social}
Stanley Wasserman and Katherine Faust.
\newblock {\em Social network analysis: Methods and applications}, volume~8.
\newblock Cambridge university press, 1994.

\bibitem{faust1997centrality}
Katherine Faust.
\newblock Centrality in affiliation networks.
\newblock {\em Social Networks}, 19(2):157 -- 191, 1997.

\bibitem{newman2010networks}
Mark Newman.
\newblock {\em Networks: an introduction}.
\newblock Oxford university press, 2010.

\bibitem{orucevic2014network}
Alma Orucevic-Alagic and Martin H{\"o}st.
\newblock Network analysis of a large scale open source project.
\newblock In {\em 40th EUROMICRO Conference on Software Engineering and
  Advanced Applications}, pages 25--29, Verona, Italy, 2014. IEEE.

\bibitem{teixeira2015lessons}
Jose Teixeira, Gregorio Robles, and Jes{\'u}s~M Gonz{\'a}lez-Barahona.
\newblock Lessons learned from applying social network analysis on an
  industrial free/libre/open source software ecosystem.
\newblock {\em Journal of Internet Services and Applications}, 6(1):1--27,
  2015.

\bibitem{damian2007collaboration}
Daniela Damian, Sabrina Marczak, and Irwin Kwan.
\newblock Collaboration patterns and the impact of distance on awareness in
  requirements-centred social networks.
\newblock In {\em International Requirements Engineering Conference}, pages
  59--68. IEEE, 2007.

\bibitem{marczak2008information}
Sabrina Marczak, Daniela Damian, Ulrike Stege, and Adrian Schroter.
\newblock Information brokers in requirement-dependency social networks.
\newblock In {\em International Requirements Engineering, 2008.}, pages 53--62.
  IEEE, 2008.

\bibitem{bhowmik2015on}
Tanmay Bhowmik, Nan Niu, Prachi Singhania, and Wentao Wang.
\newblock On the role of structural holes in requirements identification: An
  exploratory study on open-source software development.
\newblock {\em ACM Trans. Manage. Inf. Syst.}, 6(3):10:1--10:30, September
  2015.

\bibitem{linaaker2016firms}
Johan Lin{\aa}ker, Patrick Rempel, Bj{\"o}rn Regnell, and Patrick M{\"a}der.
\newblock How firms adapt and interact in open source ecosystems: Analyzing
  stakeholder influence and collaboration patterns.
\newblock In {\em Requirements Engineering: Foundation for Software Quality},
  pages 63--81. Springer, 2016.

\bibitem{johnson2008exploring}
Gerry Johnson, Kevan Scholes, and Richard Whittington.
\newblock {\em Exploring corporate strategy: text \& cases}.
\newblock Pearson Education, 2008.

\bibitem{newcombe2003client}
Robert Newcombe.
\newblock From client to project stakeholders: a stakeholder mapping approach.
\newblock {\em Construction Management and Economics}, 21(8):841--848, 2003.

\bibitem{mendelow1991stakeholder}
A~Mendelow.
\newblock Stakeholder mapping.
\newblock In {\em Proceedings of the 2nd International Conference on
  Information Systems}. MA Cambridge, 1991.

\bibitem{munir2017open}
Hussan Munir, Johan Lin{\aa}ker, Krzysztof Wnuk, Per Runeson, and Bj{\"o}rn
  Regnell.
\newblock Open innovation using open source tools: a case study at sony mobile.
\newblock {\em Empirical Software Engineering}, 23(1):186--223, Feb 2018.

\bibitem{mitchell1997toward}
Ronald~K Mitchell, Bradley~R Agle, and Donna~J Wood.
\newblock Toward a theory of stakeholder identification and salience: Defining
  the principle of who and what really counts.
\newblock {\em Academy of management review}, 22(4):853--886, 1997.

\bibitem{barnett2011encyclopedia}
George~A Barnett.
\newblock {\em Encyclopedia of social networks}.
\newblock Sage Publications, 2011.

\bibitem{damian2010requirements}
Daniela Damian, Irwin Kwan, and Sabrina Marczak.
\newblock Requirements-driven collaboration: Leveraging the invisible
  relationships between requirements and people.
\newblock In {\em Collaborative software engineering}, pages 57--76. Springer,
  2010.

\bibitem{henkel2008champions}
Joachim Henkel.
\newblock Champions of revealing-the role of open source developers in
  commercial firms.
\newblock {\em Industrial and Corporate Change}, 18(3):435--471, December 2008.

\bibitem{dahlander2006man}
Linus Dahlander and Martin~W. Wallin.
\newblock A man on the inside: Unlocking communities as complementary assets.
\newblock {\em Research Policy}, 35(8):1243 -- 1259, 2006.

\bibitem{bird2012examining}
Christian Bird and Nachiappan Nagappan.
\newblock Who? where? what?: examining distributed development in two large
  open source projects.
\newblock In {\em Proceedings of the 9th IEEE Working Conference on Mining
  Software Repositories}, pages 237--246. IEEE Press, 2012.

\bibitem{gonzalez2013understanding}
Jesus~M Gonzalez-Barahona, Daniel Izquierdo-Cortazar, Stefano Maffulli, and
  Gregorio Robles.
\newblock Understanding how companies interact with free software communities.
\newblock {\em IEEE software}, 30(5):38--45, 2013.

\bibitem{barrat2004architecture}
Alain Barrat, Marc Barthelemy, Romualdo Pastor-Satorras, and Alessandro
  Vespignani.
\newblock The architecture of complex weighted networks.
\newblock {\em Proceedings of the National Academy of Sciences of the United
  States of America}, 101(11):3747--3752, 2004.

\bibitem{opsahl2010node}
Tore Opsahl, Filip Agneessens, and John Skvoretz.
\newblock Node centrality in weighted networks: Generalizing degree and
  shortest paths.
\newblock {\em Social Networks}, 32(3):245--251, 2010.

\bibitem{freeman1978centrality}
Linton~C Freeman.
\newblock Centrality in social networks conceptual clarification.
\newblock {\em Social networks}, 1(3):215--239, 1978.

\bibitem{hanneman2005introduction}
Robert~A Hanneman and Mark Riddle.
\newblock {\em Introduction to social network methods}.
\newblock University of California Riverside, 2005.

\bibitem{brandes2001faster}
Ulrik Brandes.
\newblock A faster algorithm for betweenness centrality*.
\newblock {\em Journal of mathematical sociology}, 25(2):163--177, 2001.

\bibitem{newman2001scientific}
Mark~EJ Newman.
\newblock Scientific collaboration networks. ii. shortest paths, weighted
  networks, and centrality.
\newblock {\em Physical review E}, 64(1):016132, 2001.

\bibitem{bonacich1987power}
Phillip Bonacich.
\newblock Power and centrality: A family of measures.
\newblock {\em American Journal of Sociology}, 92(5):1170--1182, 1987.

\bibitem{runeson2012casestudy}
Per Runeson, Martin H{\"o}st, Austen Rainer, and Bj{\"o}rn Regnell.
\newblock {\em Case Study Research in Software Engineering - Guidelines and
  Examples}.
\newblock Wiley, 2012.

\bibitem{schaarschmidt2015firms}
Mario Schaarschmidt, Gianfranco Walsh, and Harald~FO von Kortzfleisch.
\newblock How do firms influence open source software communities? a framework
  and empirical analysis of different governance modes.
\newblock {\em Information and Organization}, 25(2):99--114, 2015.

\bibitem{joblin2017classifying}
Mitchell Joblin, Sven Apel, Claus Hunsen, and Wolfgang Mauerer.
\newblock Classifying developers into core and peripheral: An empirical study
  on count and network metrics.
\newblock In {\em Proceedings of the 39th International Conference on Software
  Engineering}, pages 164--174. IEEE Press, 2017.

\bibitem{bird2006mining}
Christian Bird, Alex Gourley, Prem Devanbu, Michael Gertz, and Anand
  Swaminathan.
\newblock Mining email social networks.
\newblock In {\em Proceedings of the 2006 international workshop on Mining
  software repositories}, pages 137--143. ACM, 2006.

\bibitem{hossain2006actor}
Liaquat Hossain, Andr{\`e} Wu, and Kenneth~KS Chung.
\newblock Actor centrality correlates to project based coordination.
\newblock In {\em Proceedings of the 2006 20th anniversary conference on
  Computer supported cooperative work}, pages 363--372. ACM, 2006.

\bibitem{o2007governance}
Siobh{\'a}n O'Mahony.
\newblock The governance of open source initiatives: what does it mean to be
  community managed?
\newblock {\em Journal of Management \& Governance}, 11(2):139--150, 2007.

\bibitem{shaikh2017governing}
Maha Shaikh and Ola Henfridsson.
\newblock Governing open source software through coordination processes.
\newblock {\em Information and Organization}, 27(2):116--135, 2017.

\end{thebibliography}

\end{document}